\documentclass{eptcs}

\usepackage{enumerate}
\usepackage{fancyvrb}
\usepackage{relsize}
\usepackage{tikz-reo}
\usepackage{subfig}
\usepackage{xspace}

\providecommand{\event}{PLACES 2012} 

\newcommand{\ANTLR}	{\textsc{Antlr}\xspace}
\newcommand{\API}	{\textsc{api}\xspace}
\newcommand{\APIs}	{{\API}s\xspace}
\newcommand{\CA}	{\textsc{ca}\xspace}
\newcommand{\DSL}	{\textsc{dsl}\xspace}
\newcommand{\DSLs}	{{\DSL}s\xspace}
\newcommand{\ECT}	{\textsc{Ect}\xspace}
\newcommand{\GPPL}	{\textsc{gppl}\xspace}
\newcommand{\GPPLs}	{{\GPPL}s\xspace}
\newcommand{\HTM}	{\textsc{htm}\xspace}
\newcommand{\IDE}	{\textsc{ide}\xspace}
\newcommand{\IO}	{\textsc{i/o}\xspace}
\newcommand{\XML}	{\textsc{xml}\xspace}
\newcommand{\TLP}	{\textsc{Tlp}\xspace}

\newcommand{\reodata}			{d}
\newcommand{\reonodea}			{{\textsf{A}}\xspace}
\newcommand{\reonodeb}			{{\textsf{B}}\xspace}
\newcommand{\reonodec}			{{\textsf{C}}\xspace}
\newcommand{\reonoded}			{{\textsf{D}}\xspace}
\newcommand{\reomergerfifo}		{{\textsf{MergerWithBuffer}}\xspace}
\newcommand{\reoalterfifo}		{{\textsf{AlternatorWithBuffer}}\xspace}
\newcommand{\reosequencerfifo}	{{\textsf{SequencerWithBuffer}}$_{\varphi}$\xspace}

\newcommand{\javamergerfifo}	{\texttt{MergerWithBuffer}\xspace}
\newcommand{\javaalterfifo}		{\texttt{AlternatorWithBuffer}\xspace}

\newcommand{\globe}	{$\,$\includegraphics[height=1.75ex]{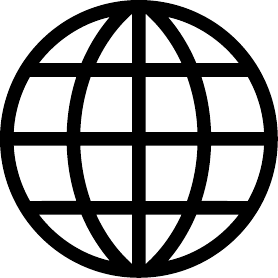}$\;$}

\newtheorem{definition}{Definition}
\newtheorem{problem}{Problem}

\begin{document}

\title{Modularizing and Specifying Protocols among Threads}
\author{%
	Sung-Shik T.Q. Jongmans
	\institute{Centrum Wiskunde \& Informatica (CWI) \\ Amsterdam, the Netherlands}
	\email{jongmans@cwi.nl}
\and
	Farhad Arbab
	\institute{Centrum Wiskunde \& Informatica (CWI) \\ Amsterdam, the Netherlands}
	\email{farhad.arbab@cwi.nl}
}

\def\titlerunning{Modularizing and Specifying Protocols among Threads}
\def\authorrunning{S.-S.T.Q. Jongmans \& F. Arbab}

\maketitle

\begin{abstract}
	We identify three problems with current techniques for implementing protocols among threads, which complicate and impair the scalability of multicore software development: implementing synchronization, implementing coordination, and modularizing protocols. To mend these deficiencies, we argue for the use of domain-specific languages (\DSL) based on existing models of concurrency. To demonstrate the feasibility of this proposal, we explain how to use the model of concurrency Reo as a high-level protocol \DSL, which offers appropriate abstractions and a natural separation of protocols and computations. We describe a Reo-to-Java compiler and illustrate its use through examples.
\end{abstract}

\raggedbottom

%
\section{Introduction}
\label{sect:intr}

With the advent of multicore processors, a new era began for many software developers of general, nonnumerical, applications: to harness the power of multicore processors, the need for writing concurrently executable code, instead of traditional sequential programs, intensified---a notoriously difficult task with the currently popular tools and technology! To alleviate the burden of implementing concurrent applications, researchers started developing new techniques for multicore programming. Examples include stream processing, transactional memory, and lock-free synchronization. However, one rather high-level aspect of multicore programming has received only little attention: the sets of rules that interacting parties must abide by when they communicate with each other---\emph{protocols}. In this paper, we investigate a new approach for implementing protocols among threads.

Many popular general-purpose programming languages (\GPPL) feature threads: concurrently executing program fragments sharing the same address space. To name a few such \GPPLs and (some of) their multithreading facilities:
\begin{itemize}
	\item Fortran has coarrays and OpenMP;
	\item C and C++ have Pthreads and OpenMP;
	\item Objective-C has Pthreads and the \texttt{NSThread} class;
	\item Visual Basic and C\# have the \texttt{System.Threading} namespace;
	\item Java has the \texttt{Thread} class.
\end{itemize}
These languages have a combined share of roughly 63\% according to the \TLP and Tiobe indexes of January 2012.%
\footnote{%
	\raggedright \globe \url{http://lang-index.sourceforge.net/}
}$^{,}$%
\footnote{%
	\raggedright \globe \url{http://www.tiobe.com/index.php/content/paperinfo/tpci/index.html}
} From these statistics, one can conclude that (a good portion of) sixty percent of software developers encounters threads regularly. Consequently, many developers will benefit from improvements to existing techniques for implementing protocols among threads. We consider such improvements not merely relevant but a sheer necessity: the current models and languages, \APIs and libraries, fail to scale when it comes to implementing protocols. Importantly, we refer to scalability not only in terms of performance but also in terms of other aspects of software development (e.g., correctness, maintainability, and reusability). Our approach in this paper takes such aspects into account.

\paragraph{Organization}

In Section~\ref{sect:prob}, we identify three problems with current techniques for implementing protocols among threads. In Section~\ref{sect:sol}, we sketch an abstract solution based on the general notion of domain-specific languages. In Section~\ref{sect:reo}, we concretize our approach for one particular such language, namely Reo. Section~\ref{sect:conc} concludes this paper.

%
\section{Problems with Implementing Protocols}
\label{sect:prob}

While threads prevail for implementing concurrency in general-purpose programming languages~(\GPPL), they also provoke controversy. Programming with threads would inflict unreasonable demands on the reasoning capabilities of software developers, partly due to the unpredictable ways in which threads interact with each other~\cite{CL00}: typically, one cannot analyze all the ways in which threads may interleave, and consequently, unforeseen---and potentially dangerous---execution paths may exist. Some propose to discard our present notion of threads, unless we improve our ways of programming with them~\cite{Lee06}. Because many existing {\GPPL}s support threads---and since this seems unlikely to change in the near future---we gear our efforts toward getting such improvements. In particular, our interest lies in solving problems with implementing \emph{protocols among threads}. In this section, we identify three such problems.

At first sight, writing the computation code and the protocol code of a program using a single language may seem only natural. Indeed, many popular \GPPLs have sufficient expressive power for doing so. Nevertheless, we consider it an inappropriate approach in many cases: typically, language designers gear \GPPLs toward implementing computations. Implementing protocols, at a suitable level of abstraction, seems a secondary concern. Consequently, these languages work well for writing computation code, but not so for developing protocol code: the low-level concurrency constructs that they provide do not coincide with the higher-level concepts needed to express protocols directly. This results in two problems that complicate ``writing code'' for protocols.

\begin{problem}
	[Implementing synchronization]
	\label{problem:sync}
	Threads communicating with each other using a shared memory, by directly reading from and writing to their common address space, must synchronize their actions. However, implementing synchronization using primitives such as locks, mutexes, sempahores, etc., comprises a tedious and error-prone activity.
\end{problem}
\begin{problem}
	[Implementing coordination]
	\label{problem:coord}
	Threads interacting with each other in structured ways, according to some protocol, require coordination to ensure that they respect this protocol. However, implementing coordination using constructs such as assignments, \textbf{if}-statements, \textbf{while}-loops, etc., produces code that only indirectly conveys a protocol, which make it a tedious and error-prone activity.
\end{problem}

Interestingly, these two problems have a common cause: the lack of \emph{appropriate abstractions} for implementing communication and interaction in {\GPPL}s. For example: software developers should specify that a thread sends two integers and receives an array of rationals for a response---not that a thread allocates shared memory and performs pointer arithmetic. Or: developers should specify that communication between two threads inhibits interaction among other threads---not that threads acquire and release locks. Or: developers should specify that threads exchange data elements synchronously (i.e., atomically)---not that threads wait on a monitor until they get notified. We believe that programming languages should enable developers implementing communication and interaction to focus on the logic of the protocols involved---not on the realization of the necessary synchronization and coordination. We (should) have compilers for that.

In addition to the two problems identified above, the lack of appropriate abstractions in {\GPPL}s causes a third problem: in the absence of proper structures to enforce (or at least encourage) good protocol programming practices, programmers frequently succumb to the temptation of not isolating protocol code. Conceptually, this problem differs from Problems~\ref{problem:sync} and~\ref{problem:coord}, because it does not complicate ``writing code'' directly. However, it does perplex essentially everything else involved in a software development process. Although notions such as ``modularization'' \cite{Par72} and ``separation of concerns'' \cite{Dij82} have long histories in computer science, linguistic support for their application in programming of concurrency protocols has scarcely received due attention. Modularization and separation of concerns have driven the development of modern programming languages and software development practices for decades. In fact, already in the early 1970s, Parnas attributed three advantages to abiding by these principles~\cite{Par72}:
\begin{quote}
	``(1) managerial---development time should be shortened because separate groups would work on each module with little need for communication; (2) product flexibility---it should be possible to make drastic changes to one module without a need to change others; (3) comprehensibility---it should be possible to study the system one module at a time.''
\end{quote}

\begin{figure}[t]
	\input{cmds-java}
	\fvset{gobble=4,tabsize=2,fontsize=\scriptsize,numbers=left,firstnumber=last,numbersep=.2cm,commandchars=\\\{\}}
	\begin{tabular}{@{} c @{} c @{}}
		\begin{minipage}{.5\textwidth}
			\begin{Verbatim}[firstnumber=1]
				\IMPORT java.util.LinkedList;
				\IMPORT java.util.concurrent.Semaphore;
				
				\PUBLIC \CLASS Main \{
					\PRIVATE LinkedList<Object> buffer;\label{fig:java:prodcons:line:prot:1}
					\PRIVATE Semaphore notEmpty;\label{fig:java:prodcons:line:prot:2}
					\PRIVATE Semaphore notFull;\label{fig:java:prodcons:line:prot:3}
					
					\PUBLIC Main() \{
						buffer = \NEW LinkedList<Object>();\label{fig:java:prodcons:line:prot:4}
						notEmpty = \NEW Semaphore(0);\label{fig:java:prodcons:line:prot:5}
						notFull = \NEW Semaphore(1);\label{fig:java:prodcons:line:prot:6}
						(\NEW Producer()).start();
						(\NEW Producer()).start();
						(\NEW Consumer()).start();
					\}
			\end{Verbatim}
		\end{minipage}
	&	\begin{minipage}{.5\textwidth}
			\begin{Verbatim}
					\PRIVATE \CLASS Producer \EXTENDS Thread \{
						\PUBLIC run() \{
							\WHILE (\TRUE) \{
								Object d = produce();
								notFull.acquire();\label{fig:java:prodcons:line:prot:7}
								buffer.offer(d);\label{fig:java:prodcons:line:prot:8}
								notEmpty.release();\label{fig:java:prodcons:line:prot:9}
					\} \} \}
					\PRIVATE \CLASS Consumer \EXTENDS Thread \{
						\PUBLIC run() \{
							\WHILE (\TRUE) \{
								notEmpty.acquire();\label{fig:java:prodcons:line:prot:10}
								Object d = buffer.poll();\label{fig:java:prodcons:line:prot:11}
								notFull.release();\label{fig:java:prodcons:line:prot:12}
								consume(d);
				\} \} \} \}
			\end{Verbatim}
		\end{minipage}
	\end{tabular}

	\caption{Java implementation of the producer--consumer example in~\cite[Algorithm 6.8]{Ben06}.}
	\label{fig:java:prodcons}
\end{figure}

\newcommand{\figjavaprodconsprotlines}	{\ref{fig:java:prodcons:line:prot:1}--\ref{fig:java:prodcons:line:prot:3},~\ref{fig:java:prodcons:line:prot:4}--\ref{fig:java:prodcons:line:prot:6},~\ref{fig:java:prodcons:line:prot:7}--\ref{fig:java:prodcons:line:prot:9}, and~\ref{fig:java:prodcons:line:prot:10}--\ref{fig:java:prodcons:line:prot:12}\xspace}

\noindent Nevertheless, popular \GPPLs do not enforce modularization of protocols. Consequently, dispersing protocol code among computation code comprises a common practice for implementing protocols among threads. To illustrate such dispersion---and its deficiencies---we discuss the producer--consumer Java code in Figure~\ref{fig:java:prodcons} (based on~\cite[Algorithm 6.8]{Ben06}). Two producer threads produce data elements and append them to a queue \texttt{buffer} (of size 1). Concurrently, a consumer thread takes elements from this queue and consumes them. While the queue \texttt{buffer} contains an element, the producers cannot append data until the consumer takes this element out of the queue. (We skip the methods \texttt{produce} and \texttt{consume}.) Easily, one can get the gist of the protocol involved in this example: the producers send---asynchro\-nously, reliably, and in arbitrary order---data elements to the consumer. In contrast, one cannot easily point to coherent segments of the source code that actually implement this protocol. Indeed, only the combination of lines \figjavaprodconsprotlines does so. In this example, thus, we have not isolated the protocol in a distinct module; we have not separated our concerns. Therefore, the ``Advantages of Modularization'' identified by Parnas do not apply. In fact, the \emph{monolithic program} in Figure~\ref{fig:java:prodcons} suffers from their opposites---the ``Disadvantages of Dispersion:''
\begin{enumerate}
	\item[($\overline{1}$)] Groups cannot work independently on computation code and protocol code of monolithic programs. Moreover, one cannot straightforwardly reuse computation code or protocol code of monolithic programs in other programs: this would require dissecting and disentangling the former.
	
	\item[($\overline{2}$)] Small changes to a protocol require nontrivial changes throughout a monolithic program. For instance, suppose that we allow the producers in the producer--consumer example to send data elements only in alternating order. Implementing such turn-taking requires significant changes.
	
	\item[($\overline{3}$)] One cannot study computation code and protocol code in isolation when they are entangled: to reason about protocol correctness, one must analyze monolithic programs in their entirety.
\end{enumerate}
The impact of these shortcomings only increases when programs grow larger, interaction among threads intensifies, and protocol complexity increases---a likely situation in the current multicore era.
\begin{problem}
	[Modularizing protocol code]
	\label{problem:modul}
	The lack of appropriate abstractions in {\GPPL}s tempts developers to disperse protocol code among code of computational tasks. In that case, developers do not isolate protocols in modules (e.g., classes, packages, namespaces), but intermix them with computations. This practice makes independently developing, maintaining, reusing, modifying, testing, and verifying protocol code problematic or impossible.
\end{problem}

\begin{figure}[t]
	\input{cmds-java}
	\fvset{gobble=4,tabsize=2,fontsize=\scriptsize,numbers=left,firstnumber=last,numbersep=.2cm,commandchars=\\\{\}}
	\begin{tabular}{@{} c @{} c @{}}
		\begin{minipage}{.5\textwidth}
			\begin{Verbatim}[firstnumber=1]
				\PUBLIC \INTERFACE Protocol \{
					\PUBLIC \VOID offer(Object o);
					\PUBLIC Object poll();
				\}
				\PUBLIC \CLASS Main \{
					\PRIVATE Protocol protocol;
					\PUBLIC Main() \{
						protocol = \NEW P();\label{fig:java:prodcons:line:p}
						(\NEW Producer()).start();
						(\NEW Producer()).start();
						(\NEW Consumer()).start();
					\}
			\end{Verbatim}
		\end{minipage}
	&	\begin{minipage}{.5\textwidth}
			\begin{Verbatim}
					\PRIVATE \CLASS Producer \EXTENDS Thread \{
						\PUBLIC run() \{
							\WHILE (\TRUE) \{
								Object d = produce();
								protocol.offer(d);
					\} \} \}
					\PRIVATE \CLASS Consumer \EXTENDS Thread \{
						\PUBLIC run() \{
							\WHILE (\TRUE) \{
								Object d = protocol.poll();
								consume(d);
				\} \} \} \}
			\end{Verbatim}
		\end{minipage}
	\end{tabular}
	\caption{Reimplementation of the producer--consumer program in Figure~\ref{fig:java:prodcons}.}
	\label{fig:java:prodcons:+}
\end{figure}

\newcommand{\figjavaprodconsXpline}	{\ref{fig:java:prodcons:line:p}\xspace}

To avoid the Disadvantages of Dispersion, we propose to isolate protocol code in separate modules. In object-oriented languages, for instance, one can achieve this by encapsulating all the protocol logic in a separate class. To illustrate this approach, we rewrote the monolithic program in Figure~\ref{fig:java:prodcons} as the \emph{modular program} in Figure~\ref{fig:java:prodcons:+}: we moved all the protocol code to a class \texttt{P} (see Section~\ref{sect:reo:dsl} for its implementation), which implements the interface \texttt{Protocol}.%
\footnote{%
	The definition of the interface \texttt{Protocol} in Figure~\ref{fig:java:prodcons:+} serves only our present discussion: not every protocol has methods \texttt{offer} and \texttt{poll}. In general, the interface of a protocol should provide methods that computation code can invoke for executing this protocol. In the context of our present discussion, \texttt{offer} and \texttt{poll} seemed appropriate names.
} To such programs, the Advantages of Modularization apply. \emph{First}, groups can develop protocol code (e.g., the implementations of the methods \texttt{offer} and \texttt{poll}) independently from computation code. Moreover, one can easily reuse protocol implementations. \emph{Second}, changing the protocol requires changing only the class implementing the protocol (e.g., the class \texttt{P}); computation code, however, remains unaffected. \emph{Third}, we can analyze the protocol in isolation by studying only the class implementing the protocol (e.g., the class \texttt{P}).
%
\section{Solution: Protocol DSLs}
\label{sect:sol}

In the previous section, we explained how the lack of appropriate abstractions complicates three aspects of implementing protocols: implementing synchronization (Problem~\ref{problem:sync}), implementing coordination (Problem~\ref{problem:coord}), and modularizing protocol code (Problem~\ref{problem:modul}). We believe that \emph{domain-specific languages}~(\DSL) offer a solution to these problems.
\begin{definition}
	[Domain-specific language~\cite{DKV00}]
	A domain-specific language is a programming language that offers, through appropriate notations and abstractions, expressive power focused on, and usually restricted to, a particular problem domain.
\end{definition}
Domain-specific languages dedicated to the implementation of protocols solve Problems~\ref{problem:sync} and~\ref{problem:coord} \emph{by this very definition}. Moreover, such \emph{protocol \DSLs} naturally force developers to isolate their protocols in modules: specifying protocols in a different language leads to a clear syntactic separation between computation code and protocol code. Consequently, using protocol \DSLs in the following workflow secures the Advantages of Modularization.
\begin{itemize}
	\item Developers write the computation code of an application in a \GPPL.
	\item Developers specify protocols among threads in a protocol \DSL.
	\item A \DSL compiler compiles protocol specifications to \GPPL code, seemlessly integrating protocols with computations.
\end{itemize}
While the benefits seem clear, one question remains: where to get these protocol \DSLs from? Must we invent them from scratch? And if so, what kinds of ``appropriate abstractions'' should they provide?

Fortunately, we do not need to design everything from the ground up: interaction and concurrency have received plenty of attention from the theoretical computer science community over the past decades, and researchers have investigated high-level models of concurrency for many years (e.g., Petri nets, process algebras). This led to various formalisms for synchronizing and coordinating parties running concurrently (e.g., actors, agents, components, services, processes, etc.). We believe that these models of concurrency provide appropriate abstractions for specifying and reasoning about protocols (albeit, not all do so equally well). In other words, the protocol \DSLs that we look for already exist. However, many of these formalisms lack sophisticated tool support, and in particular, the kind of compiler mentioned above. Therefore, we consider implementing such code generation tools among the main goals in our efforts toward alleviating the burden of programming with threads. But which existing concurrency formalism should we focus on? 

%
\section{Reo as a Protocol DSL}
\label{sect:reo}

One model of concurrency has our particular interest: Reo~\cite{Arb04,Arb11}, an interaction-based model of concurrency with a graphical syntax, originally introduced for coordinating components in component-based systems. As with other models of concurrency, Reo has a solid foundation: there exist various compositional semantics \cite{JA12} for describing the behavior of Reo programs, called \emph{connectors}, along with tools for analyzing them. This includes both functional analysis (detecting deadlock, model-checking) and reasoning about nonfunctional properties (computing quality-of-service guarantees). Its declarative nature, however, distinguishes Reo from other models of concurrency. Using Reo, software developers specify \emph{what}, \emph{when}, \emph{where}, and \emph{why} interaction takes place; not \emph{how}. Indeed, Reo does not feature primitive actions for sending or receiving data elements. Rather, Reo considers interaction protocols as constraints on such actions. In stark contrast to traditional models of concurrency, Reo's constraint-based notion of interaction has the advantage that to formulate (specify, verify, etc.) protocols, one does not need to even consider any of the alternative sequences of actions that give rise to them.

Using Reo, computational threads remain completely oblivious to protocols that compose them into, and coordinate their interactions within, a concurrent application: their code contains no concurrency primitive (e.g., semaphore operations, signals, mutex, or even direct communication as in send/receive). The sole means of communication for a computational thread consists of \IO actions that it performs on its own input/output ports. To construct an application, one composes a set of such threads together with a protocol by identifying the input/output ports of the computational threads with the appropriate output/input nodes of a Reo connector that implements the protocol. A Reo compiler then generates the proper multithreaded application code.

We proceed as follows. In Section~\ref{sect:reo:eg}, we explain the main concepts of Reo through three example connectors, each of which represents a protocol that one can use in the producer--consumer example of Section~\ref{sect:prob}. In Section~\ref{sect:reo:dsl}, we discuss our Reo-to-Java compiler.

\subsection{Reo by Example: Producer--Consumer Protocols}
\label{sect:reo:eg}

\begin{figure}
	\newcommand{\HEIGHT}{82pt}
	\begin{minipage}{.49\textwidth}
		\subfloat[Syntax.]{\label{fig:mergerfifo:syntax}%
			\vbox to \HEIGHT {%
				\vfil
				\hbox to .5\linewidth {
					\hfil \begin{tikzpicture}[baseline, node distance=1.5cm]
	\node[reonode]		(C) [label=above:\scriptsize $\enspace$ \textsf{C}] {}; 
	\node[reobnode]		(A) [above left of=C, label=left:\scriptsize \textsf{A}] {};
	\node[reobnode]		(B) [below left of=C, label=left:\scriptsize \textsf{B}] {};
	\node[reobnode]		(D) [right of=C, label=right:\scriptsize \textsf{D}] {};
	
	\draw[sync]			(A) to node {} (C);
	\draw[sync]			(B) to node {} (C);
	\draw[fifoempty]	(C) to node {} (D);
\end{tikzpicture} \hfil
				}
				\vfil
			}
		}
		\subfloat[Semantics.]{\label{fig:mergerfifo:semantics}%
			\vbox to \HEIGHT {%
				\vfil
				\hbox to .45\linewidth {
					\hfil \begin{tikzpicture}[baseline]
	\node[state, initleft] 		(Q) {};
	\node[state, node distance=2cm]	(R) [right of=Q] {};
	\path[->]	(Q)	edge [trans, out=-15, in=-165]		node [below] {\scriptsize $\{ \textsf{A} , \textsf{C} \}$}		(R);
	\path[->]	(Q)	edge [trans, out=-75, in=-105]		node [below] {\scriptsize $\{ \textsf{B} , \textsf{C} \}$}		(R);
	\path[->]	(R) edge [trans, out=165, in=15]		node [above] {\scriptsize $\{ \textsf{D} \}$}					(Q);
\end{tikzpicture} \hfil
				}
				\vfil
			}
		}
		
		\captionsetup{width=.8\textwidth}
		\caption{Syntax and semantics, as a port automaton, of \reomergerfifo.}
		\label{fig:mergerfifo}
	\end{minipage}
	\hfil
	\begin{minipage}{.49\textwidth}
		\subfloat[Syntax.]{\label{fig:alterfifo:syntax}%
			\vbox to \HEIGHT {%
				\vfil
				\hbox to .5\linewidth {
					\hfil \begin{tikzpicture}[baseline, node distance=1.5cm]
	\node[reonode]		(C) [label=above:\scriptsize $\enspace$ \textsf{C}] {}; 
	\node[reobnode]		(A) [above left of=C, label=left:\scriptsize \textsf{A}] {};
	\node[reobnode]		(B) [below left of=C, label=left:\scriptsize \textsf{B}] {};
	\node[reobnode]		(D) [right of=C, label=right:\scriptsize \textsf{D}] {};
	
	\draw[sync]			(A) to node {} (C);
	\draw[fifoempty]	(B) to node {} (C);
	\draw[syncdrain]	(A) to node {} (B);
	\draw[fifoempty]	(C) to node {} (D);
\end{tikzpicture} \hfil
				}
				\vfil
			}
		}
		\subfloat[Semantics.]{\label{fig:alterfifo:semantics}%
			\vbox to \HEIGHT {%
				\vfil
				\hbox to .45\linewidth {
					\hfil \begin{tikzpicture}[baseline]
	\node[state, initleft] 			(Q) {};
	\node[state, node distance=2cm]	(R) [right of=Q] {};
	\node[state, node distance=2cm]	(S) [below of=Q] {};
	\node[state, node distance=2cm]	(T) [right of=S] {};
	
	\path[->]	(Q)	edge [trans]		node [above] {\scriptsize $\{ \textsf{A} , \textsf{B}, \textsf{C} \}$}		(R);
	\path[->]	(R)	edge [trans]		node [ left] {\scriptsize $\{ \textsf{D} \}$}							(T);
	\path[->]	(T)	edge [trans]		node [below] {\scriptsize $\{ \textsf{C} \}$}							(S);
	\path[->]	(S) edge [trans]		node [right] {\scriptsize $\{ \textsf{D} \}$}							(Q);
\end{tikzpicture} \hfil
				}
				\vfil
			}
		}
		
		\captionsetup{width=.8\textwidth}
		\caption{Syntax and semantics, as a port automaton, of \reoalterfifo.}
		\label{fig:alterfifo}
	\end{minipage}

\end{figure}

Figures~\ref{fig:mergerfifo:syntax},~\ref{fig:alterfifo:syntax}, and~\ref{fig:sequencerfifo:syntax} show three example connectors (i.e., Reo programs): graphs of nodes and arcs, which we refer to as \emph{channels}. We refer to nodes that admit \IO operations as \emph{boundary nodes} and draw them as open circles in figures. Intuitively, one can interpret the graph representing a Reo connector as follows: data elements, dispatched on input (boundary) nodes by output operations, move along arcs to other nodes, which replicate them if they have multiple outgoing channels, along to output (boundary) nodes, from which input operations can fetch them. Groups of such (input, output, and transport) activities may take place atomically. Importantly, communicating parties that perform \IO operations on the boundary nodes of a connector remain oblivious to how the connector routes data: parties that fetch or dispatch data elements do not know where these elements come from or go to.

The connector in Figure~\ref{fig:mergerfifo:syntax} specifies the same protocol as the one embedded in the Java code in Figure~\ref{fig:java:prodcons}. We can explain the behavior of this connector, named \reomergerfifo, best by discussing the \emph{port automaton}~\cite{KC09} that describes its semantics. Figure~\ref{fig:mergerfifo:semantics} shows this automaton (derived automatically from Figure~\ref{fig:mergerfifo:syntax}): every state corresponds to an internal configuration of \reomergerfifo, while every transition describes a step of the protocol specified by \reomergerfifo. Transitions carry a \emph{synchronization constraint}: a set containing those nodes through which a data element passes in an atomic protocol step. Thus, in the initial state of \reomergerfifo, a data element passes either nodes \reonodea and \reonodec or nodes \reonodeb and \reonodec. Every element that passes \reonodec subsequently arrives at a buffer with capacity 1. We represent this buffer with a rectangle in Figure~\ref{fig:mergerfifo:syntax}. While the buffer remains full, no data elements can pass \reonodea, \reonodeb, or \reonodec. In that case, the only admissible step results in the element stored in the buffer leave the buffer and pass through \reonoded.

Figure~\ref{fig:alterfifo} shows another connector, named \reoalterfifo, that one can use in the producer--consumer example. In contrast to \reomergerfifo, \reoalterfifo forces the producers to synchronize (represented by the arrow-tailed edge between nodes \reonodea and \reonodeb): only if they can dispatch data elements simultaneously, the connector allows them to do so. In that case, the data element dispatched on \reonodea passes node \reonodec and enters the horizontal buffer; concurrently, the data element dispatched on \reonodeb enters the diagonal buffer. In the next protocol step, the data element in the horizontal buffer leaves this buffer and passes node \reonoded. Subsequently, the data element in the diagonal buffer leaves this buffer, passes \reonodec, and enters the horizontal buffer. Finally, the data element now in the horizontal buffer (originally dispatched on \reonodeb) leaves this buffer and passes \reonoded. Thus, \reoalterfifo first synchronizes the producers, and second, it offers their data elements in alternating order to the consumer.

\begin{figure}[t]
	\newcommand{\HEIGHT}{111pt}
	\hfil
	\subfloat[Syntax (with $\varphi \equiv {[}\reonodea{]} = \mbox{``foo''}$).]{\label{fig:sequencerfifo:syntax}%
		\vbox to \HEIGHT {%
			\vfil
			\hbox to .45\textwidth {
				\hfil \begin{tikzpicture}[baseline, node distance=1.5cm]
	\node[reonode]		(C) [] {}; 
	\node[reonode]		(A2) [above left of=C] {};
	\node[reonode]		(A1) [left of=A2] {};
	\node[reobnode]		(A) [left of=A1, label=left:\scriptsize \textsf{A}] {};
	\node[reonode]		(B2) [below left of=C] {};
	\node[invis]		(B1) [left of=B2] {};
	\node[reobnode]		(B) [left of=B1, label=left:\scriptsize \textsf{B}] {};
	\node[reobnode]		(D) [right of=C, label=right:\scriptsize \textsf{D}] {};

	\node[reonode, node distance=1.414cm]		(X1) [above of=B1] {};
	\node[reonode]		(X2) [right of=X1] {};
	\node[reonode, node distance=.707cm]		(X3) [below of=X2] {};
	\node[reonode]		(X4) [left of=X3] {};
	
	\draw[sync]			(A2) to node {} (C);
	\draw[sync]			(B2) to node {} (C);
	\draw[sync]			(B) to node {} (B2);
	\draw[fifoempty]	(C) to node {} (D);
	\draw[fifoempty]	(X1) to node {} (X2);
	\draw[sync]			(X2) to node {} (X3);
	\draw[fifofull]		(X3) to node {} (X4);
	\draw[sync]			(X4) to node {} (X1);
	\draw[syncdrain]	(X1) to node {} (A1);
	\draw[syncdrain]	(X3) to node {} (B2);
	\draw[sync]			(A) to node {} (A1);
	\draw[filter]		(A1) to node {} (A2);
\end{tikzpicture}  \hfil
			}
			\vfil
		}
	}
	\hfil
	\subfloat[Semantics.]{\label{fig:sequencerfifo:semantics}%
		\vbox to \HEIGHT {%
			\vfil
			\hbox to .45\textwidth {
				\hfil \begin{tikzpicture}[baseline]
	\node[state, initleft] 				(Q) {};
	\node[state, node distance=4.5cm]	(R) [right of=Q] {};
	\node[state, node distance=3cm]		(S) [below of=Q] {};
	\node[state, node distance=4.5cm]	(T) [right of=S] {};
	
	\path[->]	(Q)	edge [trans]			node [above]		{\scriptsize $\{ \textsf{A} \}$, $[\textsf{A}] = \mbox{``foo''} \wedge m' = [\textsf{A}]$}		(R);
	\path[->]	(Q)	edge [trans, sloped]	node [above left]	{\scriptsize $\{ \textsf{A} \}$, $[\textsf{A}] \neq \mbox{``foo''}$}							(T);
	\path[->]	(R)	edge [trans]			node [right]		{\scriptsize $\begin{array}{@{}c@{}} \{ \textsf{D} \} \, , \\ {}[\textsf{D}] = m \end{array}$}	(T);
	\path[->]	(T)	edge [trans]			node [below]		{\scriptsize $\{ \textsf{B} \}$, $m' = [\textsf{B}]$}											(S);
	\path[->]	(S) edge [trans]			node [left]			{\scriptsize $\begin{array}{@{}c@{}} \{ \textsf{D} \} \, , \\ {}[\textsf{D}] = m \end{array}$}	(Q);
	\path[->]	(S)	edge [trans, sloped]	node [above left]	{\scriptsize $\{ \textsf{A} \}$, $[\textsf{A}] \neq \mbox{``foo''} \enspace$}					(R);
\end{tikzpicture} \hfil
			}
			\vfil
		}
	}
	\hfil

	\caption{Syntax and semantics, as a constraint automaton, of \reosequencerfifo.}
	\label{fig:sequencerfifo}
\end{figure}

Figure~\ref{fig:sequencerfifo} shows a third connector, named \reosequencerfifo, that one can use in the producer--consumer example. The protocol specified by this connector differs in two significant ways from \reomergerfifo and \reoalterfifo. The first difference relates to (the lack of) synchronization: unlike \reoalterfifo, \reosequencerfifo does not force the producers to synchronize before they dispatch their data elements. (Similar to \reoalterfifo, however, \reosequencerfifo orders the sequence in which data elements arrive at the consumer.) The second difference relates to the \emph{data-sensitivity} that \reosequencerfifo exhibits: the zigzagged edge in Figure~\ref{fig:sequencerfifo:syntax} represents a \emph{filter channel} and we call $\varphi$ a \emph{filter constraint}: only those data elements satisfying its filter constraint propagate through a filter channel. In this example, we assume a simple filter constraint, namely $[\reonodea] = \mbox{``foo''}$, which means: the data element passing \reonodea equals the string ``foo''.%
\footnote{%
	Alternatively, one can formulate filter constraints as regular expressions or \emph{patterns}. See~\cite{Arb04}.
}
In other words, if a producer dispatches ``foo'' on \reonodea, the (right-horizontal) buffer becomes filled with ``foo''; otherwise, the filter loses the dispatched data element, which means essentially, its producer has wasted its turn. In general, filter channels facilitate the specification of protocols whose execution depends on the content of the exchanged data.

Port automata cannot express the semantics of connectors with filter channels. For that, we need a stronger formalism: \emph{constraint automata}~(\CA)~\cite{BSAR06}, which support richer transition structures than port automata. Instead of only a synchronization constraint, transitions of \CA carry also a \emph{data constraint}: an expression about what the data passing particular nodes should look like in some protocol step. Figure~\ref{fig:sequencerfifo:semantics} shows the constraint automaton that describes the semantics of \reosequencerfifo (we omitted its nonboundary nodes from this \CA). The symbols $m$ and $m'$ refer to the content of the (right-horizontal) buffer while and after a transition fires: $m' = [\reonodea]$ means that this buffer contains the value exchanged through \reonodea after a transition; $[\reonoded] = m$ means that the content of this buffer passes through \reonoded during a transition.

\subsection{Compiling Reo to Java}
\label{sect:reo:dsl}

Next, we discuss how to use Reo as a \DSL for implementing protocols: we present here an early version of our Reo-to-Java compiler, because it is simpler to explain. Although we focus on Java here, we emphasize the generality of our approach: nothing in Reo prevents us from compiling Reo to Fortran, C, C++, Objective-C, C\#, or Visual Basic.

Before we can compile anything, we need a means to implement the ``paper-and-pencil drawings'' of connectors. We use the Reo \IDE for this purpose, called \emph{the Extensible Coordination Tools} (\ECT).%
\footnote{%
	\raggedright \globe \url{http://reo.project.cwi.nl/}
} The \ECT consists of a collection of Eclipse plug-ins, including a drag-and-drop editor for drawing connector diagrams. Under the hood, the \ECT stores and manipulates such diagrams as \XML documents. These \XML documents serve as input to our Reo-to-Java compiler, detailed next.

Previously, we introduced Reo in terms of how data elements move through a connector, not unlike dataflow programming. Compiling connectors to some kind of distributed application, therefore, may seem an obvious choice: nodes naturally map to processing elements (e.g., cores), and the connections between processing elements can serve as channels. However, this approach has several shortcomings. \emph{First}, the network topology of the hardware may not correspond with the topology of the connector that we want to deploy. \emph{Second}, emulating Reo channels with hardware connections requires additional computations from the processing elements connected. This destroys the original idea of mapping Reo channels to hardware connections. \emph{Third}, achieving the global atomicity, synchronization, and exclusion emerging in a connector requires complex distributed algorithms. Such algorithms inflict communication and processing overhead, deteriorating performance. In short, construing connectors and their topology too literally seems a bad idea in the context of compilation. Instead, our Reo-to-Java compiler compiles connectors based on their constraint automaton (\CA) semantics.

The \ECT ships with the \CA of many common channels, including those in Figures~\ref{fig:mergerfifo},~\ref{fig:alterfifo}, and~\ref{fig:sequencerfifo}. By combining such primitive \CA, through the act of composition \cite{BSAR06}, the \ECT can automatically compute the \CA of larger connectors. We use this open source \CA library in our Reo-to-Java compiler: on input of an \XML document specifying a connector, our compiler first computes the corresponding \CA. Subsequently, it annotates this \CA with Java identifiers. Finally, it produces a Java class using \ANTLR's StringTemplate technology \cite{Par07}. One can use the resulting class as any Java class. By using \CA for compiling connectors, we conveniently abstract away their nonboundary nodes.

\begin{figure}[t]
	\input{cmds-java}
	\fvset{gobble=4,tabsize=2,fontsize=\scriptsize,numbers=left,firstnumber=last,numbersep=.2cm,commandchars=\\\{\}}
	\begin{tabular}{@{} c @{} c @{}}
		\begin{minipage}{.5\textwidth}
			\begin{Verbatim}[firstnumber=1]
				\PUBLIC \CLASS {\javamergerfifo} \EXTENDS Thread \{\label{fig:java:mergerfifo:line:class}
					\comment{The current state.}
					\PRIVATE State current = State.EMPTY;\label{fig:java:mergerfifo:line:current}
					\PRIVATE \ENUM State \{ FULL, EMPTY \}
					
					\comment{The boundary nodes of this connector.}
					\PRIVATE Port A; \PRIVATE Port B; \PRIVATE Port D;\label{fig:java:mergerfifo:line:ports}
					
					\comment{The data constraints this connector checks and}
					\comment{the memory cells this connector has access to.}
					\eolcomment{---snip---}

					\comment{A random number generator for selecting transitions.}
					Random random = new Random();

					\comment{Constructs a MergerWithBuffer.}
					\PUBLIC MergerWithBuffer(Port A, Port B, Port D) \{
						\THIS.A = A; \THIS.B = B; \THIS.D = D;
						
						\comment{Initialize data constraints.}
						\eolcomment{---snip---}
					\}

					\comment{Runs the state machine modeling MergerWithBuffer.}
					\PUBLIC \VOID run() \{\label{fig:java:mergerfifo:line:run:begin}
						\WHILE (\TRUE) \{\label{fig:java:mergerfifo:line:run:loop}
							\SWITCH (current) \{
							\CASE State.EMPTY:
								\SWITCH (random.nextInt(2)) \{\label{fig:java:mergerfifo:line:run:select}
								\CASE 0: tFromEmptyToFullA(); \BREAK;
								\CASE 1: tFromEmptyToFullB(); \BREAK;
								\}
								\BREAK;
							\CASE State.FULL:
								\eolcomment{---snip---}
								\BREAK;
					\} \} \}\label{fig:java:mergerfifo:line:run:end}
			\end{Verbatim}
		\end{minipage}
	&	\begin{minipage}{.5\textwidth}
			\begin{Verbatim}
					\comment{Makes a transition from state EMPTY to state FULL, firing A.}
					\PRIVATE \VOID tFromEmptyToFullA() \{\label{fig:java:mergerfifo:line:transa:begin}
						
						\comment{Lock and get pending writes.}
						Set<Write> writesOnA = A.lockAndGetWrites();\label{fig:java:mergerfifo:line:writes}
						\IF (writesOnA.isEmpty()) \{ abort(); \RETURN; \}\label{fig:java:mergerfifo:line:trans:abort:1}
						
						\comment{Lock and get pending takes.}
						Set<Take> takesOnD = A.lockAndGetTakes();\label{fig:java:mergerfifo:line:takes}
						\IF (takesOnD.isEmpty()) \{ abort(); \RETURN; \}\label{fig:java:mergerfifo:line:trans:abort:2}
						
						\comment{Check the synchronization and data constraints.}
						\IF (\comment{---snip---}) \{\label{fig:java:mergerfifo:line:trans:check}
							
							\comment{Process writes and takes.}
							A.performAndUnlock(\comment{---snip---});\label{fig:java:mergerfifo:line:performwrite}
							D.performAndUnlock(\comment{---snip---});\label{fig:java:mergerfifo:line:performtake}

							\comment{Update memory cells.}
							\eolcomment{---snip---};
							
							\comment{Update state.}
							current = State.FULL;
						\}
						abort();\label{fig:java:mergerfifo:line:trans:abort:3}
					\}\label{fig:java:mergerfifo:line:transa:end}

					\comment{Makes a transition from state EMPTY to state FULL, firing B.}
					\PRIVATE \VOID tFromEmptyToFullB() \{\label{fig:java:mergerfifo:line:transb:begin}
						\eolcomment{---snip---}
					\}\label{fig:java:mergerfifo:line:transb:end}

					\comment{Aborts a transition by unlocking all that it may have locked.}
					\PRIVATE abort() \{
						A.unlockWrites(); B.unlockWrites();\label{fig:java:mergerfifo:line:unlock:1}
						D.unlockTakes();\label{fig:java:mergerfifo:line:unlock:2}
				\} \}
			\end{Verbatim}
		\end{minipage}
	\end{tabular}
	\caption{(Parts of the) Java class generated by compiling \reomergerfifo (see also Figure~\ref{fig:mergerfifo}).}
	\label{fig:java:mergerfifo}
\end{figure}

To illustrate the compilation process, suppose that we want to compile \reomergerfifo for use in the producer--consumer example of Section~\ref{sect:prob}. After drawing \reomergerfifo using the \ECT, we feed the corresponding \XML document to our Reo-to-Java compiler. This tool automatically generates a Java class \javamergerfifo based on the \CA semantics of \reomergerfifo. More precisely, \javamergerfifo objects run state machines representing the protocol specified by \reomergerfifo. Figure~\ref{fig:java:mergerfifo} shows (parts of) the Java class generated by compiling \reomergerfifo. We discuss some of its salient aspects.

\begin{itemize}
	\item
	The class \javamergerfifo extends the class \texttt{Thread} (line~\ref{fig:java:mergerfifo:line:class}). By running connectors in their own thread, we enable them to proactively sense their environment for \IO operations; with massive-scale concurrent hardware, ample cores to run connectors on should always exist.

	\item Instances of \javamergerfifo listen to three \emph{ports} (line~\ref{fig:java:mergerfifo:line:ports}), which grant ``computation threads'' access to the boundary nodes of \reomergerfifo. All interaction between computation threads and a \javamergerfifo object occurs through the latter's ports: computation threads can perform \IO operations---writes and takes---on ports, which in turn suspend threads until their operations succeed. More technically, ports extend concurrent data structures called \emph{synchronization points}:%
	\footnote{%
		Synchronization points resemble $\pi$-calculus channels.
	} pairs of sets---one containing pending writes, another containing pending takes---supporting and admissible to \emph{two-phase locking} schemes (see below) \cite{BHG87}.

	The class \texttt{SyncPoint} exposes the following methods:
	\begin{itemize}
		\item \texttt{lockAndGetWrites()} locks and returns the set of pending write operations (line~\ref{fig:java:mergerfifo:line:writes}).
		\item \texttt{lockAndGetTakes()} locks and returns the set of pending take operations (line~\ref{fig:java:mergerfifo:line:takes}).
		\item \texttt{unlockWrites()} and \texttt{unlockTakes()} unlock the sets of writes and takes (lines~\ref{fig:java:mergerfifo:line:unlock:1}--\ref{fig:java:mergerfifo:line:unlock:2}).
		\item \texttt{performAndUnlock(Write)} performs the specified write operation (first parameter) and unlocks the set of pending write operations (line~\ref{fig:java:mergerfifo:line:performwrite}).
		\item \texttt{performAndUnlock(Take,Object)} performs the specified take operation (first parameter) by passing it the data element to take (second parameter) and unlocks the set of pending take operations (line~\ref{fig:java:mergerfifo:line:performtake}).
	\end{itemize}

	\item 
	The overridden method \texttt{run()} implements a state machine corresponding to the input \CA of the compilation process (lines~\ref{fig:java:mergerfifo:line:run:begin}--\ref{fig:java:mergerfifo:line:run:end}). The main loop never terminates (line~\ref{fig:java:mergerfifo:line:run:loop}). In each iteration, it randomly selects a transition (line~\ref{fig:java:mergerfifo:line:run:select}) going out of the current state (line~\ref{fig:java:mergerfifo:line:current}). We collect code responsible for making transitions in separate methods (lines~\ref{fig:java:mergerfifo:line:transa:begin}--\ref{fig:java:mergerfifo:line:transa:end} and~\ref{fig:java:mergerfifo:line:transb:begin}--\ref{fig:java:mergerfifo:line:transb:end}).
	
	\item 
	An important step in the process of making a transition consists of checking its synchronization and data constraints (line~\ref{fig:java:mergerfifo:line:trans:check}). To do this in a thread-safe manner, a \javamergerfifo object employs a two-phase locking scheme. During the \emph{growing phase}, it acquires the locks of:
	\begin{itemize}
		\item the set of pending writes of each port providing access to an input node (line~\ref{fig:java:mergerfifo:line:writes});
		\item the set of pending takes of each port providing access to an output node (line~\ref{fig:java:mergerfifo:line:takes}).
	\end{itemize}
	A \javamergerfifo object locks only the sets of those boundary nodes that actually occur in the constraints under investigation. Later, during the \emph{shrinking phase}, it releases these locks again (lines~\ref{fig:java:mergerfifo:line:trans:abort:1},~\ref{fig:java:mergerfifo:line:trans:abort:2}, and~\ref{fig:java:mergerfifo:line:trans:abort:3}). Only between phases, a \javamergerfifo object checks the constraints under investigation. If they hold, it fires the corresponding transition, transporting data elements accordingly and removing the operations involved from the sets it has locked (lines~\ref{fig:java:mergerfifo:line:performwrite}--\ref{fig:java:mergerfifo:line:performtake}).
\end{itemize}

\begin{figure}[t]
	\input{cmds-java}
	\fvset{gobble=4,tabsize=2,fontsize=\scriptsize,numbers=left,firstnumber=last,numbersep=.2cm,commandchars=\\\{\}}
	\begin{tabular}{@{} c @{} c @{}}
		\begin{minipage}{.5\textwidth}
			\begin{Verbatim}[firstnumber=1]
				\PUBLIC \CLASS P \IMPLEMENTS Protocol \{
					\PRIVATE Port A = new Port();
					\PRIVATE Port B = new Port();
					\PRIVATE Port D = new Port();
					\PRIVATE Map<Thread, Port> threads = 
							\NEW ConcurrentHashMap<Thread, Port>();
					
					\PUBLIC P() \{
						\NEW {\javamergerfifo}(A, B, D).start();\label{fig:java:p:line:mergerfifo}
					\}
			\end{Verbatim}
		\end{minipage}
	&	\begin{minipage}{.5\textwidth}
			\begin{Verbatim}
					\PUBLIC Object poll() \{ \RETURN D.take(); \}\label{fig:java:p:line:take}
					\PUBLIC \VOID offer(Object o) \{
						Thread thread = Thread.currentThread();
						\IF (!threads.containsKey(thread))
							\SYNCHRONIZED (\THIS) \{
								threads.put(thread, 
									!threads.containsValue(A) ? A : B);
							\}
						threads.get(thread).write(o);\label{fig:java:p:line:write}
				\} \}
			\end{Verbatim}
		\end{minipage}
	\end{tabular}
	\caption{Class \texttt{P}.}
	\label{fig:java:p}
\end{figure}

\newcommand{\figjavapwriteline}			{\ref{fig:java:p:line:write}\xspace}
\newcommand{\figjavaptakeline}			{\ref{fig:java:p:line:take}\xspace}
\newcommand{\figjavapmergerfifoline}	{\ref{fig:java:p:line:mergerfifo}\xspace}

\noindent To use the class \javamergerfifo in the producer--consumer example of Section~\ref{sect:prob}, we should incorporate it in the implementation of the class \texttt{P}, encountered before on line \figjavaprodconsXpline in Figure~\ref{fig:java:prodcons:+}; Figure~\ref{fig:java:p} shows this implementation. Line \figjavapwriteline specifies that a producer performs a (blocking) write operation on the port assigned to it; line \figjavaptakeline specifies that a consumer performs a (blocking) take operation. The rest of \texttt{P} consists of initialization code. The latter characterizes the use of Reo as a protocol \DSL: implementations of the \texttt{Protocol} interface serve as wrappers for compiled connectors, encapsulating all the protocol logic.

To change Figure~\ref{fig:java:prodcons} such that it respects the protocol specified by \reoalterfifo requires nontrivial modifications across the source code. In contrast, we can straightforwardly implement a class \texttt{Q} \textbf{implements} \texttt{Protocol} and replace \texttt{P()} with \texttt{Q()} on line \figjavaprodconsXpline in Figure~\ref{fig:java:prodcons:+}. In fact, \texttt{Q} would differ from \texttt{P} only on line \figjavapmergerfifoline in Figure~\ref{fig:java:p}: in \texttt{Q}, we would construct an \javaalterfifo object instead of a \javamergerfifo object. Similarly, we can use the protocol specified by \reosequencerfifo. This shows that using Reo, we can easily change protocols without affecting computation code.%
\footnote{%
	More precisely, handwritten computation code and protocol code generated by our Reo compiler communicate only through shared ports; these ports do not change when replacing one connector with another. Thus, unless the number of ports changes, a syntactically valid program remains syntactically valid.
}

This subsection demonstrates the feasibility of modularizing protocols and implementing protocol {\DSL}s. We remark that this approach does not preclude the use of \emph{dedicated implementations} for certain parts of a protocol. For instance, consider the buffer of \reomergerfifo. Our Reo-to-Java compiler implements this buffer using shared memory and explicit locks (transparent to software developers using Reo, though). But suppose that the architecture we deploy our producer--consumer program on features also \emph{hardware transactional memory} (\HTM) \cite{HM93}. Our approach allows one to write a dedicated implementation of buffers that exploits this \HTM. Subsequently, we can replace the standard buffer implementation with this dedicated implementation.\footnote{%
	Roughly, first, we write an \HTM-based implementation of buffers. Second, we remove the connection between nodes \reonodec and \reonoded in Figure~\ref{fig:alterfifo:syntax}. Third, we generate code for the subconnector containing nodes \reonodea, \reonodeb, and \reonodec and for the subconnector containing node \reonoded. (The latter consists of only node \reonoded. However, by the semantics of Reo, we can replace this by an equivalent connector of two synchronizing nodes.) Thus, we now have two compiled connectors. Finally, we place the \HTM-based implementation of buffers between these two compiled connectors, as an active entity: our dedicated buffer implementation, which performs writes and takes on \reonodec and \reonoded, runs in its own thread---effectively, it operates as a fourth party involved in the protocol. Interestingly, the ``real'' communicating parties remain oblivious to the introduction of this fourth party. Further optimization can eliminate the active entity that performs writes and takes on C and D, merging its functionality in the behavior of nodes \reonodec and \reonoded.
} Thus, besides high-level constructs by default, our approach offers developers the flexibility of applying lower-level languages if necessary.

%
\section{Concluding Remarks}
\label{sect:conc}

Our current work focuses on improving our Reo-to-Java compiler. For instance, the classes currently generated by our compiler execute sequentially. We can parallelize this rather straightforwardly by checking ports for appropriate \IO operations concurrently for different transitions. However, we speculate that we can get even better performance if, instead, we optimize at the semantic level: we wish to decompose automata into ``smaller'' automata that can execute concurrently without synchronization while preserving the original semantics (see \cite{JCP12} for preliminary results). Another potential optimization involves scheduling: the formal semantics of connectors provide very tangible information for scheduling threads. Exploiting this information should yield substantial performance gains. Hopefully, such improvements make our approach a competitive alternative to lower level approaches also in terms of performance.

In recent years, \emph{session types} \cite{HVK98,HYC08} have entered the realm of object-oriented programming (recent work includes \cite{CCD+09,DDMY09,GVR+10,HMB+11,HKP+10}). Although session types comprise a valuable new technique for programming with threads, we wonder if the abstractions provided by them suffice. Still, we consider it a very interesting development, especially since Reo does not feature types; extending Reo with session types would comprise a significant improvement.

Finally, although we focused on implementing protocols among threads in this paper, the Reo-to-Java compiler presented has proven itself useful also in the domain of Web Service orchestration \cite{JSS+12}.

\bibliographystyle{eptcs}
\bibliography{main}

\end{document}